# Gateway Controller with Deep Sensing: Learning to be Autonomic in Intelligent Internet of Things

Rahim Rahmani and Ramin Firouzi

Department of Computer and Systems Sciences, University of Stockholm, Sweden

{rahim, ramin@dsv.su.se}

*Abstract*—The Internet of Things(IoT) will revolutionize the Future Internet through ubiquitous sensing. One of the challenges of having the hundreds of billions of devices that are estimated to be deployed would be rise of an enormous amount of data, along with the devices ability to manage. This paper presents an approach as a controller solution and designed specifically for autonomous management, connectivity and data interoperability in an IoT gateway. The approach supports distributed IoT nodes with both management and data interoperability with other cloud-based solutions. The concept further allows gateways to easily collect and process interoperability of data from IoT devices. We demonstrated the feasibility of the approach and evaluate its advantages regarding deep sensing and autonomous enabled gateway as an edge computational intelligence.

*Index Terms*—Intelligent IoT, Open Distributed system, Autonomous Gateway, Context-aware pervasive system, Deep Sensing, Advanced Machine Learning, Edge computing, Fog computing.

## I. INTRODUCTION

The Internet of Things (IoT) technologies have changed several aspects, such as reducing the possibilities of failures and delays, enhancing productivity and efficiency, improving real-time decision making, solving critical problems, and creating new and innovative experiences in the e.g. energy industry. In the coming years, with constantly improving technology and increasing integration of large-scale data, even small- and medium-scale players in the e.g. energy industry would be drawn to adopt IoT solution and services. This, along with identification, data capture, and processing capabilities will let the IoT provide data and services to be used by future applications. which combines network technologies with wireless computing, IoT capability and some subjects related to Artificial Intelligence (AI), is to create an environment where the connectivity of devices is embedded in such a way that the connectivity is unobtrusive yet useful, and always available. Advanced in the Intelligent Internet of Things (IIoT) is expected to drive the future controller in IoT gateway and gaining attention both from academia and the industry. It is envisioned that IIoT will impact almost every aspect of life and hundreds of billions of things are believed to be connected by near future.

With the increasing number of connected things, a number of key features and challenges arise. Device heterogeneity is necessary to allow devices using a variety of protocols and architectures to coexist in the same networks and function correctly when interconnected [1]. IoT networks need to be scalable to handle the increasing number of connected devices. Some important issues that these scalable IoT networks need to be able to handle are: naming, addressing, communication, networking, information, knowledge management, service provisioning and management [1],[2]. One method that seeks to solve above mentioned features and challenges, and enable an IoT infrastructure, is the usage of rule-based cloud computing applications and services or multimodal reasoning. These applications, while convenient and certainly useful, may not be efficient enough for future IoT applications, especially when scaled to service the large numbers of devices that are expected to connect to the internet of things. A wide range of IoT devices sensing and computing applications require time-series measurements to generate inputs for various parameters estimations and classification applications. For Deep Sensing oriented problems such as activity and context recognition a typical approach is to compute appropriate features derived from raw sensor data and that directly addresses the challenges in the division between knowledge-based and behavior-based which AI has been fundamental to achieving successful applications with in the field of autonomous [3]. However up to now this division has had few repercussions for reinforcement learning in the autonomous IoT gateways controller. The main objective of deploying reinforcement-learning methods in the IoT gateway controller is to establish a correct mapping from a set of abstraction observation to a set of high level actions [4]. Algorithm developed within this general framework can be used in different fields without any modification. For each particular service the definition of the sets of states and actions is task of the Deep Coder [5,6].

Therefore, we focus on a controller solution design specifically for autonomous management, connectivity and data interoperability in IoT gateway are critical in order to realize an autonomous enabled gateway as an edge computational intelligence. In this paper we propose an autonomous controller concept for management and connectivity decision making in distributed IoT devices. The concept allows gateways to collect and process interoperability of data from IoT devices easily. To achieve this, we propose new strategies that make our solution more scalable. The contribution of this paper can be summarized as follows. We present the design and implementation details of our proposed controller solution support autonomy and scalability. Our solution support both management Interoperability and data Interoperability with other cloud-based solutions. The architecture is scalable and promotes ease-of-use.

The rest of the paper is structured as follows. Section 2 defines the background and motivation of Gateway with deep sensing learning to be Autonomic and Section 3 presents the related works and section 4 describes the conceptual model of the framework; Section 5 presents the performance evaluation and the conclusions are provided in section 6.

## II. BACKGROUND AND MOTIVATION

In this section we briefly discuss the background and our motivation behind this work.

### A. Gateway with Deep Sensing Learning to be Autonomic

Properties of future IIoT, correspond to a classification layers as seen from Figure1. This classification can be compared with the vision of autonomic computing loop [7]. The autonomic self-execute programs in order to make an autonomous gateway solution specifically for autonomous cooperative decision-making in massively Distributed IoT networks. Execution of algorithms on autonomous gateway depends on the progression on collective IoT devices and enabled program reasoner (rules) in order to facilitate humanity. This can also be seen from AWS IoT's rule-engines where rules are added whenever required and further in XpertRule [8]. Today we face that IoT and Artificial Intelligence (AI) thereby, Machine Learning (ML) will be inseparable. The reason is that up until now IoT only focused on collecting and sharing data, it did not focus on providing insight to the data. The existing approaches of deploying ML for IoT gateway are heavily cloud-based which increased response time. The need for two-level intelligence was highlighted in [9,10,11]. Providing intelligence at edge implies reaping value from the collected raw-data by the gateway, i.e. an IoT controller. This paper is an extension of [7,9] and one of several working towards a common goal of implementing the autonomous IoT gateway controller. Therefore, we focus on the management Interoperability and data Interoperability with other cloud-based solutions. The main research question to answer is what the most suitable architecture and deep learning algorithms are for an efficiency and versatility on the edge gateways controller for synthesizing programs within the devices management and data interoperability to achieve deep sensing and autonomous gateway as an edge computational intelligence.

Due to such difficulties in one side and other side to enable elasticity for dynamic load adaption and simplified resource management need to utilize an autonomic edge / gateway as shown in Figure 1. One of key terms suggested as solution related to massive adoption is orchestration and slicing of resources and services [7]. But selecting of the slices almost without a dynamic load adaption of unstructured devices are note possible.

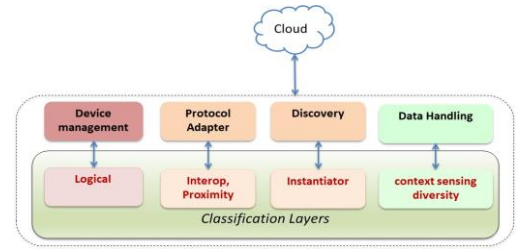

Figure 1 Gateway Learning to be autonomic

## III. RELATED WORKS

In this section we briefly discuss the motivation behind this work and related works and comprehensive introduction to the development and the stat-of-the-art of the gateway controller intelligence. As important of many different platforms and actors in the growing IoT industry and unfortunately the autonomic gateway in Internet of Things (IoT) field is unable to flourish to its full potential toward Internet of Every things (IoE). In a future where everything is connected to everything, isolated applications with lack of standardization will end up in an inefficient [12].

In [13] the authors introduce implementation and comparison of two fuzzy-based systems for IoT device selection in opportunistic networks. The approach is focusing only of device selection. Since edge computing with deep learning it is promising solution to combine edge and AI together. This new paradigm of intelligent is called intelligence-edge AI [14]. In [15] the authors address the challenges in Deep Neural Networks (DNN) with in device-edge. The approach is focusing on DNN co-inference framework. The [16] approaches three existing architectures of mobile intelligence in detail. Edge intelligence refers to a set of connected systems and devices for data collection, processing and analysis in close to where data is collected, to purposing the speed of the data processing and data interoperability. The [17] discuss key issue of extending Deep Learning (DL) from the cloud to the edge of the network under the multiple constraints of networking, communication, computing power, and energy consumption and the authors introduce how to develop edge computing architecture to achieve the performance of DL training and inference. The [18] discuss a seamlessly blending machine learning approach into the design and operation of mobile/embedded systems. The user could customize intelligent applications [19] by training DL models with self-generated data. The demo for achieves a better performance shows by verification on wearable devices [20] for multimodal deep learning under wearable data and for convolutional layers within the architecture. The edge intelligent [21] which uses Federated Learning (FL) to collaboratively train the typing prediction model on mobile devices. The FL [21] is distributed machine learning approach which enables training on a large corpus of decentralized data residing. The [22] mainly focus on the architecture and applications of federation learning. The [23] focus on how to realize the training and inference of DL models on a mobile device. The [24] provide an overview on using a DL, to facilitate the analytics and learning in the IoT domain. They articulating IoT data characteristics and identifying two major treatments for IoT data from a machine learning perspective, IoT big data analytics and IoT streaming data analytics. The survey [25] focus on how DL introduce essential background in DL techniques with applications to networking. They discuss several techniques and platforms that facilitate the efficient deployment of DL onto mobile systems. The [26] mainly focus on inter-availability between edge computing and DL. The authors discuss optimization problems at edge with DL approaches and applying DL in the context of edge computing. By contrast in this paper we pay more attention to the connected autonomous gateway at the edge to support both management interoperability and data interoperability with other cloud-based solutions. To allow gateways to collect and process interoperability of data from IoT devices easily we focus on how to realise the controller intelligence and there are two key components in to adapt to arbitrary protocols in order to subscribe to various kinds of devices. The complete process of implementing Deep Coder should involve data collection, device management and model training and inference. Hence the scope of our approach involves how to autonomously manage devices and how to integrating cloud and edge computing for data interoperability.

*A. Autonomous Device Management*

Today there are several big companies retailing cloud device management solutions. The common thing for them all is that the connected devices are defined by the end-user and require manual configuration. Neat solution for configuration seems to be one of the selling arguments when advertising the solution [27,28,29]. The traditional approach relies on manual configuration which consume to much bandwidth and will therefore be unsuitable in a ubiquitous communication network. It is clear that the device management needs to be autonomous. The "Large-Scale autonomic IoT Gateway (LSG)", from a recent study in [30] automatically discover and connect to arbitrary devices. The LSG solution is suitable on constrained hardware, which is a step in the direction of IoT. The solution is more similar to a preprogrammed interface supporting the most common devices of today, than to the proposed fully-generic gateway in our approach which it able to produce own code and by that change its own behavior.

The smart city would require a device manager that automatically discover nearby sensor population and if containing devices of interest group and store the devices based on the context they belong to when a new sensor population is discovered in the network its content (sensor data) needs to be classified and assigned one or several contexts with a service for every sensor to handle continuous communication. These services are called Sensor Agents (SA). Since edge computing [31] has recently been proposed as extension of cloud computing to push services to the access network. In scenarios where the edge gateway will perform bulk operations upon larger groups of devices such procedure could be costly. Instead of treating the devices on individual level the device could be arranged as context based logical network where the publish/subscribe technique is used [32]. Since a gateway needs to be able to change its usage mode at any given moment, it should not persist all of the information that it gathers, it should only persist the information which is currently meaningful. The device manager should keep its storage free from outdated contexts and retrieve contexts and their related SA as they become active again. Currently there is at least four different protocols used by devices [33] and the number will grow [34].

An autonomous gateway needs to be able to adapt to arbitrary protocols in order to subscribe to various kinds of devices. A protocol adapter is therefore necessary.

*B. Domain-Specific Language (DSL)*

A domain-specific language is a programming language or executable specification language that is restricted to a certain domain, by explicitly covering the requirements for the domain [35]. Common examples of DSLs are HTML and SQL as they are only used in one domain each the web domain respectively the database domain. In contrast to general-purpose languages a DSL usual have a more restricted program space which gives the main reason to why our approach makes use of a DSL a smaller search space would be less time consuming when synthesizing a program [34,35,36].

*C. Deep Coder*

Due to the demands on efficiency and versatility on the edge the process of synthesizing programs within the device management need to be as fast as possible. The computational power of future gateway will on the other hand be constrained. In recent studies the use of machine-learning techniques within applications for constrained smart devices has been proven both accurate and efficient [37,38]. This paper shows the possibility of using advanced machine learning to optimize the software on the future gateway.

*D. Integrating Cloud and Edge computing*

Recently, specialized IoT controllers with distributed intelligence capabilities have been proposed as a solution to implementing future IoT networks. These controllers combine the benefits of both cloud and edge computing by utilizing AI and machine learning [9]. The controller, having access to input data from all connected things can use that raw data to contextualize, analyze, and extract information for use in decision making and management of said things [9]. The use of AI-based distributed intelligence along with machine learning and belief networks allowed the controller to make predictions based on earlier beliefs and thus handle uncertain situations [9]. Using a controller like the one proposed [9] is a step towards a functional future IoT solution. This paper will examine the idea that an IoT gateway and its controller such as described above improved upon by the integration of deep learning and inductive program synthesis (IPS).

*E. Work related to our concept and approach*

All of the above-mentioned technologies are either directly or indirectly necessary for the development and maintaining of enabling distributed intelligence controller such as the one proposed in this paper. We will examine the idea that an IoT gateway or controller such as described above may be improved upon by the integration of deep learning and inductive program synthesis (IPS). The proposed autonomic gateway for IoT will make use of Deep Coder's [5,6] LIPS approach as a source of inspiration to create an autonomic system that combines the strengths of cloud and edge computing for use in future IoT networks. In order to learning in the approach Q-learning implement as the learning part of LIPS and very simple, customized, DSL's was designed as the basis for synthesizing programs within the separate modules of the artefact. This approach was chosen, as opposed to, for example, implementing Deep Coder [5,6] directly, in order to allow DSL functions specialized to the modules of the artefact to be used that are either not included in or incompatible with Deep Coder [5,6].

## IV. PROPOSED APPROACH

This section describes modelling of the proposed approach of the paper. We first describe the workflow of the model and follows with the proposed solution's algorithms.

### A. Work Flow

The paper proposes to provide an entire gateway to counter the influx of context information in the IoT domain by providing intelligence both at the edge and cloud as illustrated in Figure 2. Edge gateway implies that intelligence based on raw-data collected by the IoT controller from the Wireless Sensor Network(WSN) be provided as fast as possible. The first task of the proposed approach is to contextualize the collected raw-data by classification layer. The contextualized data will be forwarded to the next layers and modules logic, interoperability, and context sensing diversity. The original design for the gateway by [7], as shown in Figure 2, proposes an autonomous gateway concept but offers no implementation of the different components. This paper is one of several working towards a common goal of implementing the entire gateway.

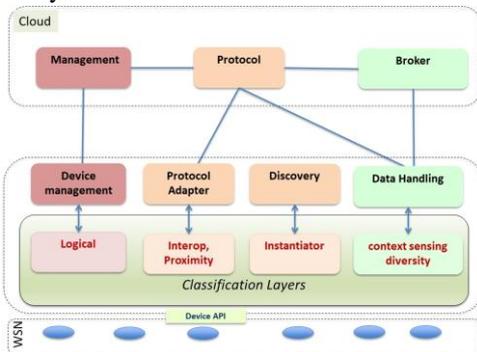
Figure 2 Gateway with deep sensing learning [7]

### B. The Requirements

The future of IoT demands a generic gateway to support arbitrary applications autonomously. The problem stated in this paper is limited to three sections shown in Figure 3. The functionalities required for the artifact will be split into several minor programs. The programs will not be isolated within the scope of the artifact, but instead will be highly coupled to the surrounding layers of a possible autonomous gateways, the cloud and Wireless Sensor Networks (WSN). The architecture of the artefact will therefore depend on assumptions made both about how the cloud and the rest of the gateway will work. Based on a given root cause the Autonomic gateway need to support the following functionalities:

1. *Automatically discover new devices*: In an IoT network, there is a need for better device discovery support to ease the deployment of machine-to-machine connectivity. Manually connecting them would be infeasible, and therefore the automatic discovery of new devices would be a core part of the gateway of the future.
2. *Provide a connection between a sensor Agent (SA) and a device:* As the main purpose of the SA is to be a logical representation of a device, it needs to be able to communicate with said device. There are a number of possible protocols used by different devices, and the SA therefore needs some assistance to communicate with the devices in the " -language-" that the device speaks.
3. *Efficiently find the proper protocol for SA:*
    The capacity of a gateway is assumed to be limited, therefor the gateway has to be economical with its computations and find the correct protocol in the most efficient manner, while also being as fast as possible.
4. *Change the way to structure devices autonomously.* The fundamental requirement of the future gateway is genericity one software solution to fit all purposes. Within the scope of this paper, the problem with genericity lies within determining which of the device properties should group them. Consider a gateway for a smart home: it would possible be preferable to manage devices by type. Another gateway used to monitor devices in a smart city would possible to manage devices by districts. The Switching between properties to cluster around need to be done autonomously when the need arises.
5. *Group the devices by classification and store them in association with the contexts they form.* The devices need to be managed in a logical and accessible way as shown in Figure 2 would need a grouping of devices that is meaningful when producing contexts. The contexts will also have to be store together with reference to the devices that created it for future use by the gateway or by request from a third party.
6. *Convert between XML and JSON.* The cloud will both backup data and forward remote application requests e.g the gateway in an autonomous vehicle could possible make use of remote traffic control or emergency services which will be connected through a service broker in the cloud. Due to differentiation of application data formats conversion will be required.

The instantiation gateway modules shown in Figure 3 further demonstrates the workflow of the proposed approach to alleviate intelligence of Gateway modules by harvesting information of things at the edge.

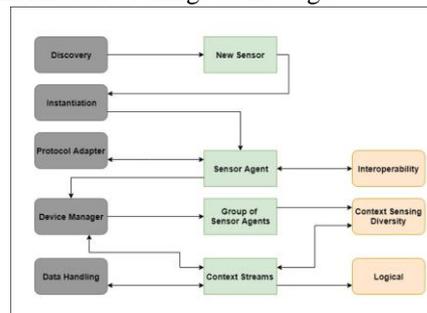
Figure 3. Outline of Workflow

Due to the limited capacities of the devices intended to run this service, a simple Q-learning approach using inductive program synthesis was selected as an implementation strategy in place of more sophisticated machine learning technology. Inductive program synthesis limits the search space and is therefore less demanding in regards to memory allocation and processing power when compared to generating programs from a bigger search space.

The design process for modules shown in Figure 3 will be described in the following subsections.

### C. Interoperability

The interoperability module enables communication between devices using different applications with the same communications protocol. Specifically, the interoperability module as designed enables translation of messages from applications running the MQTT protocol; gmqtt [39], Eclipse Paho Python [40,42], and the standard MQTT format.

### D. Logical

The logical module of the gateway can make simple decisions based on information from devices and sensors in the network as well as previous experience. The design in this paper can handle two modalities of information for logical reasoning, such as learning the relationship between the values of two sensors and using that information to control an actuator.

### E. Context Sensing Diversity

Context sensing diversity takes in a group of contexts along with new sensors and uses IPS to decide in which of those contexts the new sensors should be placed, if any. The conditions for deciding matches in contexts include proximity, i.e. physical, temporal, or social, and other similarities between the different values of the (SA).

### F. Protocol Adapter

The following requirements have been identified: A) Provide a connection between a SA and a device, B) Efficiently find the proper protocol for SA, C) Locate protocol adapter and its functionality, D) Retrieved data by protocol adapter, E) Retrieved dynamic sensor data by SA as soon as a sensor will be included in the WSN, F) Select the feasible protocol by protocol adapter and, G) Rank the protocol. The above mentioned requirements will be described in the following subsections.

#### 1) Assess and Select

Creating an instance for each SA that listens on the network would overconsume memory computation power. To avoid a static protocol with thousands of SAs for updating we designed a dynamic sink connection via a Broker. Instances of the protocol adapter connected to different Broker could associate the devices in the sink with the corresponding SAs. The consequences of making all of the SAs retrieve data from the Protocol Adapter (PA) is the need to be running constantly to update and thus consuming a lot of computing power. Since the commonly used protocols in IoT applications such as CoAP and MQTT protocols are publish/ subscribe oriented (e.g MQTT "subscribes to topic and CoAP "observes" resources) to different streams of data and it would be feasible to automatically route the data to the corresponding SA. A threads based solution would be required in case of the both hardware and software gateway limitations.

In our approach we selected a protocol ranking algorithm based on statistics as shown in Figure 4.

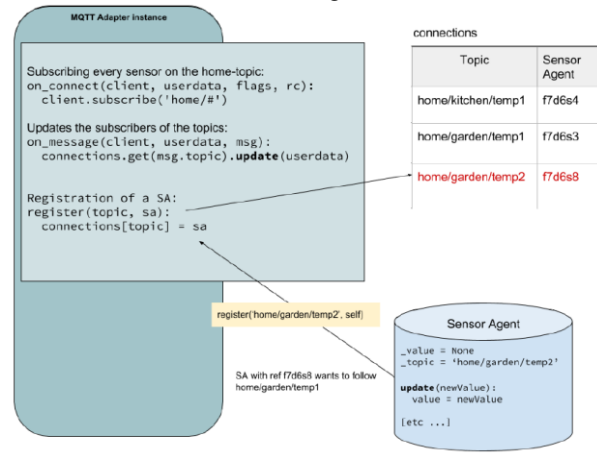

Figure 4. The Protocol Adapter data flow

#### 2) The Architecture

The architecture consists of a MQTT [39,42] adapter class a CoAP [40,43] adapter class and a main program (main PA). The adapter classes will register SAs in a hash table which maps the resources identifier (topic or URI) to the corresponding ID of an SA. Receiving messages to the adapter will, for instance from the broker, the correct SA will be pulled out from the table by using the resource identifier that the message came with and given the received message as shown in Figure 4. Discovery will delegate the task of newly arriving new sensor devices of a broker to the adapter classes as shown in Figure 3. To be able to manage that an unmatched topics or URIs will be sent to the Classification Layer for instantiation instead of being discarded as shown in Figure 5. The architecture allows adding further adapter classes that to be instantiated from the main Protocol Adapter(PA). The main PA is looking for suitable protocols to instantiate the proper adapter to connect with an incoming broker. The main PA makes use of a list that keeps track of the usage of each adapter-class. The list contains pointers to respective adapter class's constructor making it possible for the main PA to loop through the list and try connecting each adapter in the order of usage.

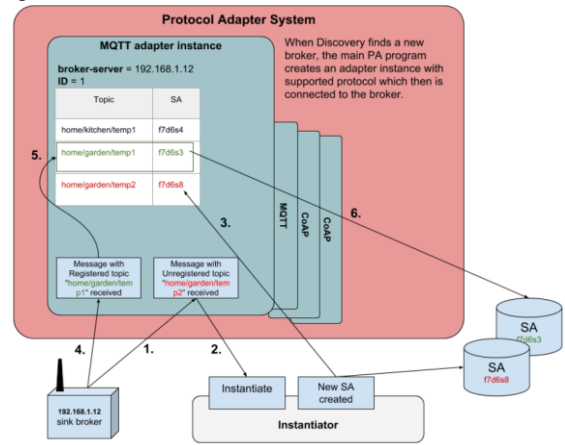

Figure 5. The protocol adapter architecture

The process of registration and forwarding data to SAs which shown in Figure 5 will be in the following steps:

A) The broker sends a message with an unknown topic or URI to the adapter instance.

B) The adapter instance tells the instantiator to create a SA.
C) The ID of the SA is sent back to the adapter instance for registration.
D) A message of a registered topic or URI is received.
E) The correct SA ID is found based on the topic or URI of the message.
F) The message is delivered to the SA with the ID from step E.

The MQTT adapter was implemented in Python using the paho mqtt client library [39] and the CoAP adapter was implemented using the CoAPthon library [40]. The main PA uses the terminal as a means of simulating all inputs and outputs within the gateway and the protocol libraries as interfaces to the brokers.

*G. Discovery*

Discovery of newly arriving devices will require an automatic discover functionality with the following consideration:
A) Devices will be connected to gateway via a sink for simplifying the data forwarding.
B) Automatic connection of the gateway to ensure maximum connection to all nearby sinks to get access to their devices.
C) Notify indirect connected devices should be send to the sink.
D) Sink communication though the proper protocol.
E) Support arbitrary protocol.
F) Utilize PA for the communication
G) Integrate with PA.

Based on these requirements the following subsections will explain the deployment of the discovery properties.

*1) Assess and Select*

Implementing several programs supporting the same future to meet the requirements D and E caused redundancy in communication network and therefore one was selected.

*2) New Sensor in Sink*

In a scenario where a new device is added to connected sink the device will start broadcasting messages. The easiest way to assign a SA to a new single device is by doing it directly when the messages is received by the PA the requirement G will be selected.

*3) New Broker*

The autonomous gateway should not connect to every available brokers it should limit the connections based on its current interests. The discovery of brokers could theoretically be implemented within the same program as PA, or as a separate component that sends the IP-addresses of the brokers to the current PA-system.

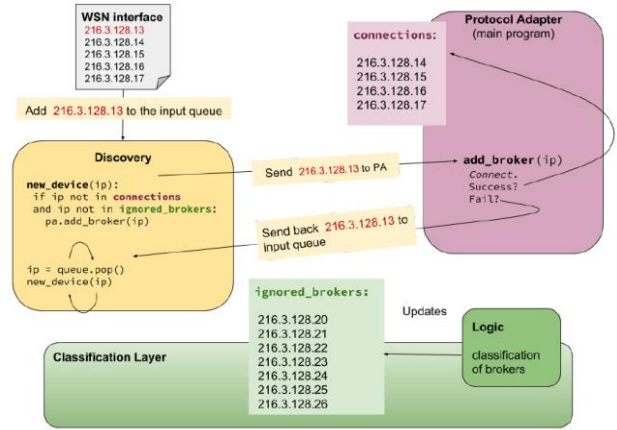

Figure 6. Discovery of new brokers

To substitute the discovery of new brokers, their IP-addresses are manually entered into the PA system, causing the same effect as an implementation shown in Figure 6.

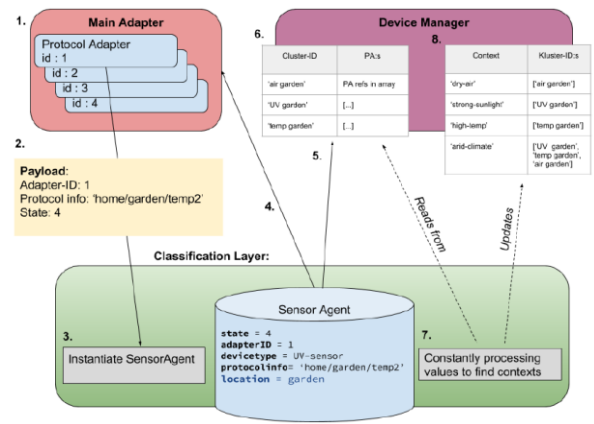

Figure 7. Discovery and Device Manager

Figure 7 depicts the discovery process in steps 1 to 6 and the device management process in steps 7 to 8.

All the steps will be explained below:
1. An adapter instance receives a message with no subscriber.
2. A message with the data from the device and a reference to the adapter instance is sent to the Classification Layer.
3. After authentication a SA is created. The location of the device is describe based on information e.g from the resource identifier of a device, etc.
4. The SA is sent back to the adapter instance with the corresponding adapter ID to be mapped with the resource identifier.
5. The SA is also sent to the Device Manager.
6. A cluster-ID is determined by the induced code of Deep Coder.
7. If not found, the cluster-ID is added to the table. The SA is added as a value associated to the cluster-ID.
8. The Classification Layer continuously processes values from clusters or primary context that contributed to that context.

Note that the first three contexts in the table are primary contexts and the fourth is a secondary context that appeared when reasoning about the first three.

*H. Device Manager*

Management of devices will be able to A) change of autonomously properties B) cluster of devices by classification C) Association of the devices to contexts.

Deep Coder [5,6] was proposed to classify SAs in dynamic way. There are two options for managing the SAs: HashMaps and a database. Both should be associated to the cluster(s) and contexts.

*1) Autonomous Clustering*

Deep Coder [4,5] is used for autonomous clustering. Deep Coder is explained in Section III.C and a program with ability to learn how to write programs which is exactly what is needed. Deep Coder could be trained to classify devices by various properties using input-output examples and it is the appropriate choice for an autonomous device management.

*2) Structure of Devices*

It is not certain that a context is determined at the time when a new device is placed within a cluster but it rather it could be assumed that other components within Classification Layer aggregates data from the clusters to find context in a constant flow. Context should be easily added or removed as they appear or lose their meaning.

We designed a Device Manager (DM) program that use Deep Coder [5,6] in a minimal way to ensure its reliability. It was modified by adding a new function to DSL, which was integrated to the system the same way as the other DSL functions, by adding it to the evaluator the dataset generator and finally training a NN to work with the new function. The new function REST removes the first element of a list. This enables the program to select an arbitrary element by iterating with the REST function. No other changes were made on the DC implementation.

*3) Clustering Program*

Deep Coder [5,6] generates a clustering program (DSL code in text format) from a set of I/O examples. The clustering program is saved to a file and further loaded into the DM. The DM reads the clustering program into a parse tree, that is evaluated by using Deep Coder [5,6] utilities to make the program executable. The DM program gets a stream of SAs as input and it then uses the clustering program to classify which cluster each of the SAs belong to (by assigning cluster ID) as shown in figure 8. The SAs are stored by the DM program together with their correct cluster.

*I. Data Handling for Interoperability*

Managing data handling requires converting between XML and JSON. We designed a Data Handler to build a program that takes the file as an argument and simply converts it to the opposite file format out of the two possible (XML, JSON). To meet the hierarchical structure of both XML and JSON we designed a tree structure which use recursion to simplify the code and avoid maximum depth constraints. To meet the Python program's non-recursive requirement we added a module with depth constraints.

*1) File Converting between XML and JSON*

Implementation of a generic extension of file converting will results in a file converter that happens to know XML and JSON rather than an explicit file converter between the two formats. The program could be easily extended to support further file formats in the future which motivates using two arguments as input. Python has limited stack depth and does not use tail call optimization and is therefore not optimized for recursive calls. Too many nested recursions could lead to run-time errors or be computationally costly. However with a test it is found that stack limit in Python is 1000, which would take a tree with 1000 levels to exceed. It could be assumed that no XML or JSON-files will be as deep as 1000 levels such as deep trees would probably be a result of an error in case of an endless loop in a server. The program file will be translated between XML and JSON according to the Table 1. The JSON object representing a XML attribute gets a prefix and the JSON object representing XML text a # prefix. Replacing the " " with an "-" and "#" with a "- -" is also common, why the converter from XML to JSON need to be able to interpret both cases.

Table 1. The JSON-XML dictionary

| JSON | XML |
|---|---|
| "e": null | <e/> |
| "e": "text" | <e>text </e> |
| "e": {"@name": "value"} | <e>text </e><e name="value" /> |
| "e": { "@name": "value", "#text": "text" } | <e name="value">text</e> |
| "e": { "a": "text", "b" : "text" } | <e><a>text</a><b>text</b></e> |
| "e": { "a": ["text", "text"] } | <e><a>text</a><a>text</a></e> |
| "e": { "#text": "text", "a": "text" } | <e>text <a>text</a></e> |

The converter is a Python file ran from the command line. The two arguments with the file name and the desired output format will be checked when the program starts. The converter will then launch either the "xml_to_json" or "json_to_xml" module. The python libraries "json" and "xml.etree. ElmentTree" are used for parsing, validating and the final printing. During the development various input was successively extended to large trees to ensure that the converters handle depth correctly. The DH converter most XML and JSON documents given to the opposite format but is unable to handle entities. The handling of entities is deemed to be excessive functionality as the devices are assumed to only send data in a simple format.

*J. Selected Programming Languages*

The repository in [5] provides and shows implementation of Deep Coder used and written it in Python and C++ as core. Due to the Deep Coder evaluator being coded in C++ made sense to implement the DM in the same language. Python is expressive as it supports object oriented as well as functional programming. The MQTT library from Eclips Paho [35] and CoAP [37] Library on github were assessed as adequate Python libraries for the implementation of the PA, both being available via the Python package manager pip. Other beneficial characteristics include concession among others

V. PERFORMANCE EVALUATION

In this section we evaluate the performance of the proposed system architecture through experiments and scenarios. Using an ex post evaluation, we employed the approach in several experiments to determine if the system lives up to our defined requirements, both functional and non-functional (latency and reliability) and thereby whether it answers the question "How can we efficiently and quickly select protocol, device management and interoperability in different application, platform and networks for the need of a Wireless Sensor Networks (WSN)? Two main experiments were performed, each experiment consisted of three sub-scenarios. The data was collected from the system and then we performed a statistical analysis of this data. Next the following case studies evaluate the validity of the approach and whether it fulfils the requirements. Both the machine learning section and the functional section of each module were tested 30 times each to achieve valid average performances and standard deviation from the results. To test the requirement of being faster than cloud computing a latency test was performed where a remote server was pinged to calculate the absolute minimum time required to forward data to the cloud.

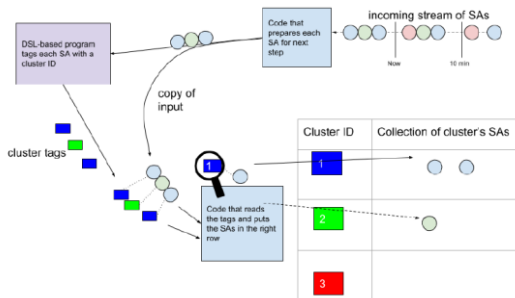

Figure 8. Clustering Devices using DeepCoder

### A. Interoperability

In this section we present results from two categories of experiments. Our experiments show that the interoperability will be improved by translating the messages of two applications, gmqtt and Eclipse Paho Python, both using the MQTT protocol. In Section A.1 we will illustrate the learning ability of the approach by demonstrating a strong kind of generation ability across the scenario. In Section A.2 will illustrate the search and run programs ability with different overloading messages.

#### 1) Learning

This experiment includes two scenarios: one scenario in which we update rules for translating messages using several iterations, while the second scenario demonstrates the ability across programs to create different messages translating formats.

*a) Scenario 1*

The first learning-scenario sought to create a rule for translating Eclipse Paho Python publish messages into a format compatible with gmqtt. The input/output example used 30 iterations returned program (2(extract_packet),3 (pack_properties)) with an average runtime of 0,166 seconds and a standard deviation of approximately 0,00173. The remaining five iterations failed to synthesize any relevant program. Those five had a similar average runtime of 0,1676 seconds and standard deviation of approximately 0,0016733.

*b) Scenario 2*

The second learning-scenario was meant to create a program for translating a message from a format used by gmqtt into an Eclipse Paho Python-compatible format. The input/output example used and the iterations return the simple program (4 (label packet)) and it had an average runtime of 0,251 seconds with standard deviation of approximately 0,006216. The exception, like those in scenario 1, failed to synthesize a relevant program at a runtime of 0,25 seconds.

#### 2) Search and Run Programs

This experiment includes two scenarios: one in which searching to find from the previously learned iterations. While a second scenario demonstrates ability across program running different messages translating formats.

*a) Scenario 1*

The interoperability module correctly found the previously learned in sections A.1.a and A.1.b programs in all 30 test iterations. The average runtime for finding and applying the program to the input data was 5,61 microseconds. with an approximate standard deviation of 0,136. The resulting output of the program was ('PUBLISH', False, 1, False, 4, 11, 'test/paho/1', 9012, (1, ('property1', 'property2', 'property3', 'property4'), ('Payload part 1', 'Payload part 2'))), matching the output format defined in learning scenario 1.

*b) Scenario 2*

The second scenario, translating a gmqtt formatted message, returned the program in all cases with a average runtime of 3,95 microseconds, slightly faster than scenario 1. The standard deviation on the other hand was higher at approximately 1,1089. The resulting output was {'command': 'PUBLISH', 'qos': 1, 'pos': 0, 'mid': 3456, 'info': (2, 'property1', 'property2'), 'packet': ('PUBLISH', False, 1, False, 7, 12, 'test/gmqtt/1', 3456, {'payload part 1': 123, 'payload part 2': 456}), 'to_process': 9}, again matching the desired output format defined in the learning section.

### B. Logic

The scenarios tested for the logic module concerns the controlling of home temperature based on dual-modal sensor values. The user who has previously stated their preferred indoor temperature is on their way home. The synthesized program calculates the time it would take to heat up the home in relation to the user's distance and activates or deactivates the actuators necessary to have the temperature be at the desired level as close to the user's arrival as possible.

#### 1) Learning

This experiment included two scenarios: one scenario which sought program synthesis that ensured the heating systems were turned on while the second scenario demonstrated the ability and across program will be created for the temperature adjustment.

*a) Scenario 1*

The first scenario sought the synthesize a program that would ensure that the heater would be turned on in time for the temperature to rise to 21°C by the time the position value of the 'phone' device reached 0. All 30 iterations succeeded in synthesizing the correct program of ((1), 0.004, (1000, '>'), (21, '<')) where (1) represents the setting of the heater actuator to its turned on state, 0.004 represents the relationship between the sensor values 1000 & 0 and 17 & 21, and (1000, '>') and (21, '<') represents key values and expected directionalities of change used for reasoning. The average runtime of the 30 iterations was approximately 0,4607 seconds with a standard deviation of approximately 0,06612.

*b) Scenario 2*

The second scenario was set up in a similar way to the first one but instead of starting a heater to raise the temperature it is designed to synthesize a program that will use a cooler to lower the temperature instead. The input/output example used is displayed in Figure 10. 28 of the 30 iterations successfully synthesized the correct program ((3), 0.004, (1000, '>'), (21, '>')). This, compared to the first scenario, used a different actuator and directionality for the goal value. The average runtime of the successful iterations was approximately 0,5207 seconds with the approximate standard deviation 0,04. The unsuccessful attempts had an average runtime of 0,499 seconds and the standard deviation was approximately 0,0344.

*2) Search and Run Program*

This experiment included three scenarios: in Scenario 1 we run the program for learning the values oscillation while in second and third scenarios demonstrating ability across programs for relationship based proximity.

*a) Scenario 1*

In the first scenario the temperature is below the desired level, but the user is too far away for the program to warrant any action. Both programs from the previous learning scenarios, ((1), 0.004, (1000, '>'), (21, '<')) and ((3), 0.004, (1000, '>'), (21, '>')), were found and ran since they are saved to the same identifying device- and sensor ids as used again here, phone, pos and temp, living_room. The resulting output was {'heater': False, 'cooler': False}, meaning both actuators remained turned off. Average runtime was 0,49 microseconds with a standard deviation of approximately 0,0414.

*b) Scenario 2*

In the second scenario the user has begun getting closer to home and the relationship between the position and the temperature is deemed sufficient to turn on the actuator connected to the heater. Like Scenario 1 both programs are found and ran, resulting in the output {'heater': True, 'cooler': False} with an average runtime of approximately 0,51567 microseconds and standard deviation of approximately 0,110662.

*c) Scenario 3*

In the last scenario the user is already at home and the temperature has risen above the desired temperature. Again, both programs described above are found and ran, this time resulting in {'heater': False, 'cooler': True}, keeping the heater turned off and turning on the cooler. The average runtime was approximately 0,52067 microseconds and the standard deviation was approximately 0,0405.

*C. Context Sensing Diversity: Adding Sensors to Relevant Contexts*

The scenarios for testing the context sensing diversity module are designed to test the functionality of synthesizing and finding programs to correctly estimate which contexts within a group of clusters are fitting for the addition of a new sensor or device.

*1) Learning*

This experiment included two scenarios: one scenario in testing program synthesizing for identifying location-based contexts than adding new sensors while in the second scenario tested program synthesizing for adding new sensors to relevant time-based contexts.

*a) Scenario 1*

The first learning scenario is to synthesize a program that can identify a location-based context and add a new sensor to it if relevant. The input example contains a new sensor, sensor101. The output example consists of the context c1 whose identifying value, 'loc', matches that of sensor101. 10 out of 30 iterations returned a correct program using the least amount of functions possible, (4 (exclude_dates), 5 (exclude_outside_std), 7(add_sensor)) and had an average runtime of approximately 0,609824 seconds and standard deviation of approximately 0,053765. Another 19 iterations resulted in functional programs that gave the correct output, albeit using more functions than necessary. Those other 19 iterations had an average timing of approximately 0,59095 seconds.

*b) Scenario 2*

The second scenario sought to synthesize a program for adding a sensor to relevant time-based contexts. The input example was the same sensor 101 as in scenario 1 and the output example is the context c3. All 30 iterations of scenario 2 successfully synthesized the most effective program for the wanted result: (1 (exclude_strings), 5 (exclude_outside_std), 7 (add_sensor)). The average runtime was approximately 0,5984 seconds with a standard deviation of 0,066.

*2) Search and Run Program*

This experiment included two scenarios. Both scenarios were based of the learning scenarios on the learning subsection above.

*a) Scenario 1*

The first scenario found and ran both programs from the learning scenarios above, since the context group and sensor labels remained the same. The resulting output was {'c1': ('loc', {'loc': (2, 0.0, 'Kista'), 'temp': (2, 3.049999999999999, 23.35), 'time': (2, 18000.0, datetime.datetime(2018, 5, 20, 5, 0))}, ('sensor1', 'sensor101'))}, showing the correct context and its updated count, standard deviation, and average value fields. The average runtime was 10,79 microseconds with a standard deviation of approximately 0,2797.

*b) Scenario 2*

The second scenario also found and ran both of the programs explained above, correctly returning {'c3': ('time', {'loc': (3, 2.309401076758503, 'Kista'), 'time': (3, 0.0, datetime.datetime(2018, 5, 20, 10, 0)), 'temp': (3, 2.734755727462488, 22.133333333333336)}, ('sensor1', 'sensor2', 'sensor101'))}. The average time was 11,44 microseconds and the standard deviation was approximately 0,279182.

*D. Connectivity and Latency*

To measure the minimum required time for online communication the public Google DNS server at address 8.8.8.8 was pinged 30 times from the same machine that ran the other tests. The average response time for the 30 pings was approximately 0,0106333 seconds, or 10633,3 microseconds. The standard deviation was approximately 0,00205918 seconds or 2059,18 microseconds.

*E. Protocol Adapter*

The MQTT adapter was connected with a number of public brokers [41] and Eclispse Msquitto [42] running on the same computer. The COAP adapter was connected with the coap.me[43] server and the server from CoAP Pthon [40] running on the same computer when no other servers could be found. The MQTT client used in the MQTT adapter was able to connect via host name, but the CoAP client was not able to connect due to IP-address allocation to a connected broker.

*1) Connecting Successfully*

To measure the connectivity 100 tests were made for each protocols. The average value for connection establishing was 0.3 s for MQTT and 0.05 s for CoAP. The mean value for both protocols was 0.18 s.

*2) Failing Connection*

The mean time was calculated by failing connection with CoAP and MQTT 100 times each. The failed connections were provoked by giving the adapter an IP for the opposite protocol. CoAP failing connection time is 0.2s on average and MQTT failing connection time is 0.5s.

*3) Ranking of the Protocols*

The average time for failing a connection can be used to illustrate how the ranking system would work in reality. The mean time for failing a connection was 0.35s which means that for every step down in the ranking system the appropriate protocol is located another 0.35s is added to the total time of searching. The time to connect to a protocol with the rank of 10 would therefore approximately take 3.5s.

*4) Random Connection Failures*

During the initial tests connections by the adapter failed randomly. In order to get statistics over the rate of failing 1000 tests were conducted for both the CoAP and MQTT adapter. The results showed that 0.5% was failing on the CoAP adapter and 2.5% failed on the MQTT adapter. The failure was spread out over the duration of the CoAP session and a mix of spread out single failures and clustered failures in the MQTT session.

*5) The CoAP Protocol Experiments*

To find possible optimizations of the CoAP-adapter two experiments as shown in Table 2. In the first the timeout (the effort time for connectivity) was set to 0.5s. In the experiment the timeout decreased to 0.1s with a second attempt following the first failing attempt.

Table 2. CoAP Experiments

| Experiment | Timeout | Failing first attempt | Failing completely |
|---|---|---|---|
| 1 | 0.5 | 0 | 0 |
| 2 | 0.1 (or 0.1 * 2) | 5 | 1 |

Failing the second attempt would mean that the adapter has failed completely to connect.

Both of the experiments were executed a 1000 times and the results in table 2 shows 0,5% failed one and only 0,1% failed connecting completely in experiment two.

*6) The MQTT Protocol Experiment*

Two experiments similar to the CoAP experiments were made for MQTT as shown in Table 3 but with the timeout set to 0.8 for the first and 0.5s for the second experiment.

Table 3. MQTT Experiments

| Experiment | Timeout | Failing first attempt | Failing completely |
|---|---|---|---|
| 1 | 0.8 | 0 | 0 |
| 2 | 0.5 (or 0.5 * 2) | 24 | 24 |

As shown in Table 3 the first experiment succeeded with no fails.

*F. Device Manager*

The most crucial parts of the Device Manager (DM) is to add incoming Sensor Agents (SA) to a cluster and training new clustering programs. It is important that the DM works fast in order to manage devices in real-time. Testing the performance (in term of speed) for adding SAs and training a new clustering program is therefore motivated. The test cases do not cover the time it takes for the DM to load the clustering program and making it executable. Testing load performance was not motivated since loading happens rarely.

*1) Adding Sensor Agents*

Inserting a SA in to DM involves running the SA through the clustering program to the tag of the SA with the cluster that it belongs to and then inserting the SA into the cluster data structure. The following two test that were repeated a 100 times each:

I. Inserting on SA into a DM keeping 10 types of SAs with 10 instances of each type.
II. Inserting one SA into a DM keeping 1000 types of SAs with 10 instances of each type.

As shown in Table 4 test cases time were 93773 ns on average longer than test case one.

Table 4. Cluster Sensor Agent

| Test case | Time in nanoseconds |
|---|---|
| 1 | 196393 |
| 2 | 290165 |

*2) Generating a New Clustering Program*

Two test cases were created to test the classification-program generation of the DM. The generated programs were expected to take one of the parameters of the SAs and return it as its cluster ID as shown in figures 9 and 10.

```
[
    {"input":[[1,10,20,5,12]],"output":1},
    {"input":[[2,11,20,5,7]],"output":2},
    {"input":[[3,12,20,9,12]],"output":3},
    {"input":[[4,13,20,9,10]],"output":4},
    {"input":[[9,13,20,9,10]],"output":9},
    {"input":[[9,8,7,6,5]],"output":9}
]
```

Figure 9. Test case one

Test case one as shown in Figure 9 classifies the SA by device type which is the first element of the SA and test case two as shown in Figure 10 classify by device ID, which is the second element of the SA.

```
[
    {"input":[[10,1,20,5,12]],"output":1},
    {"input":[[11,2,20,5,7]],"output":2},
    {"input":[[12,3,20,9,12]],"output":3},
    {"input":[[13,4,20,9,10]],"output":4},
    {"input":[[13,9,20,9,10]],"output":9},
    {"input":[[8,9,7,6,5]],"output":9}
]
```

Figure 10. Test case two

The program generation was executed 50 times and the times for generation was recorded. Table 5 shows the average time execution for each of the test cases.

Table 5. Time to generate classification code

| I/O example set | Time in seconds |
|---|---|
| Test case one | 3.95 |
| Test case two | 7.11 |

*G. The Data Handling*

Five examples with different characteristics were created to test the Data Handling (DH). A normal XML-depth is usually around two-three Depth which is why most of the examples were around that depth. Two of the examples were made unrealistic with a depth of 10 to put some pressure on the DH.

Table 6. Characteristics and results from XML to JSON

| Test | Depth | Lines | Objects | Time |
|---|---|---|---|---|
| 2-lvl, short | 2 | 18 | 13 | 13 |
| 2-lvl, mid | 2 | 123 | 100 | 59 |
| 3-lvl, long | 3 | 183 | 140 | 100 |
| 10-lvl, short | 10 | 24 | 10 | 154 |
| 10-lvl, mid | 10 | 51 | 24 | 408 |

*1) XML to JSON*

As shown in the table 6 a XML file with a depth of ten level distributed over 51 lines was the input causing the longest execution time when converting from XML to JSON.

Table 7. Spearman correlation test (XML to JSON)

| | | Depth | Lines | Objects | Time |
|---|---|---|---|---|---|
| Depth | Correlation Coefficient | 1.000 | .000 | -.316 | .949* |
| | Sig. (2-tailed) | . | 1.000 | .604 | .014 |
| | N | 5 | 5 | 5 | 5 |
| Lines | Correlation Coefficient | .000 | 1.000 | .900* | .200 |
| | Sig. (2-tailed) | 1.000 | . | .037 | .747 |
| | N | 5 | 5 | 5 | 5 |
| Objects | Correlation Coefficient | -.316 | .900* | 1.000 | -.100 |
| | Sig. (2-tailed) | .604 | .037 | . | .873 |
| | N | 5 | 5 | 5 | 5 |
| Time | Correlation Coefficient | .949* | .200 | -.100 | 1.000 |
| | Sig. (2-tailed) | .014 | .747 | .873 | . |
| | N | 5 | 5 | 5 | 5 |

*. Correlation is significant at the 0.05 level (2-tailed).

As shown in the table 7 the results from a Spearman correlation test show that there is a strong and significant correlation between consumed time and the depth.

*2) JSON to XML*

As shown in table 8 a JSON file with a depth of three levels distributed over 304 lines was the input causing the longest execution time when converting from JSON to XML.

Table 8. Characteristics and results from JSON to XML

| Test | Depth | Lines | Objects | Time |
|---|---|---|---|---|
| 2-lvl, short | 2 | 32 | 5 | 34 |
| 3-lvl, mid | 3 | 206 | 40 | 47 |
| 3-lvl, long | 3 | 304 | 60 | 58 |
| 10-lvl, short | 10 | 25 | 9 | 14 |
| 10-lvl, mid | 10 | 57 | 21 | 18 |

The results of the Spearman correlation test shown in the table 9 that there is a strong and significant correlation between time and amount of lines.

Table 9. Spearman correlation test (JSON to XML)

| | | Depth | Lines | Objects | Time |
|---|---|---|---|---|---|
| Depth | Correlation Coefficient | 1.000 | -.264 | -.105 | -.632 |
| | Sig. (2-tailed) | . | .668 | .668 | .252 |
| | N | 5 | 5 | 5 | 5 |
| Lines | Correlation Coefficient | -.264 | 1.000 | .900* | .900* |
| | Sig. (2-tailed) | .668 | . | .037 | .037 |
| | N | 5 | 5 | 5 | 5 |
| Objects | Correlation Coefficient | .105 | .900* | 1.000 | .700 |
| | Sig. (2-tailed) | .866 | .037 | . | .188 |
| | N | 5 | 5 | 5 | 5 |
| Time | Correlation Coefficient | -.632 | .900* | .700 | 1.000 |
| | Sig. (2-tailed) | .252 | .037 | .188 | . |
| | N | 5 | 5 | 5 | 5 |

*. Correlation is significant at the 0.05 level (2-tailed)

## VI. CONCLUSION

A major problem for IoT gateways has been performance on management, connectivity and interoperability. Most IoT gateway controllers are between one to two orders of magnitude of device management and devices connectivity slower than cloud-based controllers, leading to both lower maximum throughput and increased energy consumption. Previous work on autonomous IoT Gateway by [30] improved management performance, but still a significant overhead remains, and the tradeoff is that the resulting of only automatically devices discovering and limitation on an autonomously management and data interoperability.

We presented a complete set of techniques to mitigate autonomously management and data interoperability on IoT gateway controller which realized an autonomous enabled gateway as an edge computational intelligence. The concept effectiveness was evaluated using a set of scenarios were elaborated for performance measurement. The results shown in section V the concept allows gateways to collect and process interoperability of data from IoT devices easily. Our solution support both management Interoperability and data Interoperability with other cloud-based solutions. A more general question is what the most suitable architecture and deep learning algorithms are for an efficiency and versatility on the edge gateways controller for synthesizing programs within the devices management and data interoperability since the computational power of the edge gateways will be constrained. In the contributions, we presented a number modification to the autonomous protocols and devices management and data interoperability on IoT gateway controller by deploying Deep Coder [5,6].